\documentclass[letterpaper, 10 pt, conference]{ieeeconf}  

\IEEEoverridecommandlockouts                              

\usepackage{cite}
\usepackage{graphicx}
\usepackage{amsmath,amssymb,amsfonts}
\usepackage{algorithmic}
\usepackage{graphicx}
\usepackage{algorithm,algorithmic}
\usepackage{hyperref}
\hypersetup{hidelinks=true}
\usepackage{makecell}
\usepackage{array}
\newcolumntype{P}[1]{>{\centering\arraybackslash}p{#1}}
\usepackage{textcomp}

\usepackage{subcaption}
\title{\LARGE \bf
Heartbeat Detection from Ballistocardiogram using Transformer Network}

\author{Ruhan Yi$^{1}$,  Mihail Popescu$^{2}$,  James M. Keller$^{1}$, Grant Scott$^{1}$, Laurel Despins$^{3}$, David Heise $^{4}$, and Marjorie Skubic$^{1}$
\thanks{$^{1}$Ruhan Yi, James M. Keller, Grant Scott, and Marjorie Skubic are with the Department of Electrical Engineering and Computer Science, University of Missouri, Columbia, MO 65211 USA
         {\tt\small ry793@mail.missouri.edu} 
    {\tt\small\{kellerj, scottgs, skubicm\}@mail.missouri.edu}}%
\thanks{$^{2}$Mihail Popescu is with the Department of Biomedical Informatics, Biostatistics and Medical Epidemiology, University of Missouri School of Medicine, Columbia, MO 65211 USA
        {\tt\small popescum@health.missouri.edu}}%
\thanks{$^{3}$Laurel Despins was with Sinclair School of Nursing, University of Missouri, Columbia, MO 65211 USA}
\thanks{$^{4}$David Heise is with the Department of Science, Technology, and Mathematics, Lincoln University, Jefferson City, MO 65101 USA
        {\tt\small HeiseD@lincolnu.edu}}
}

\begin{document}

\maketitle
\thispagestyle{empty}
\pagestyle{empty}

\begin{abstract}

Longitudinal monitoring of heart rate (HR) and heart rate variability (HRV) can aid in tracking cardiovascular diseases (CVDs), sleep quality, sleep disorders, and reflect autonomic nervous system activity, stress levels, and overall well-being. These metrics are valuable in both clinical and everyday settings. In this paper, we present a transformer network aimed primarily at detecting the precise timing of heart beats from predicted electrocardiogram (ECG), derived from input Ballistocardiogram (BCG). We compared the performance of segment and subject models across three datasets: a lab dataset with 46 young subjects, an elder dataset with 28 elderly adults, and a combined dataset. The segment model demonstrated superior performance, with correlation coefficients of 0.97 for HR and mean heart beat interval (MHBI) when compared to ground truth. This non-invasive method offers significant potential for long-term, in home HR and HRV monitoring, aiding in the early indication and prevention of cardiovascular issues.

\end{abstract}

\section{INTRODUCTION}

Cardiovascular diseases (CVDs) are the leading cause of mortality globally, contributing substantially to both deaths and disabilities. In 2021 alone, CVDs accounted for 20.5 million deaths, constituting approximately one-third of all global deaths\cite{lindstrom2022global}. The significant impact of cardiovascular health on mortality and morbidity highlights the importance of monitoring heart health. Effective monitoring can help in the early detection, prevention, and management of CVDs, which can lead to improved quality of life and a reduction in healthcare costs\cite{oude2023health}. Monitoring cardiovascular health can be approached through various methods, among which tracking heart rate (HR) and heart rate variability (HRV) are particularly prominent. These parameters offer valuable insights into the autonomic regulation of the heart and can serve as early indicators of potential cardiovascular issues, including hypertension, arrhythmias, and heart failure 
\cite{thayer2010relationship}\cite{soares2014physical}. By continuously tracking HR and HRV, healthcare professionals can identify abnormal patterns allowing for timely interventions. Beyond cardiovascular health, these metrics are also essential for sleep stage monitoring, providing insights into sleep quality and identifying disorders such as sleep apnea\cite{stein2012heart}\cite{tobaldini2013heart}. Thus, integrating HR and HRV monitoring into routine healthcare practices is essential for enhancing cardiovascular health and reducing the global burden of CVDs.

There are various methods for monitoring HR, each with its advantages and limitations. The electrocardiogram (ECG) is considered the gold standard for HR monitoring due to its high accuracy and detailed heart activity data. ECG is extensively used in clinical settings, but it requires electrodes to be attached to the body, which can be inconvenient for continuous or long-term monitoring outside of a clinical environment. Methods that have been proposed to estimate HR from ECG are mostly based on the detection of R-peaks. The most well-known algorithm was developed by Pan and Tompkins\cite{pan1985real}. There exist other methods using wavelet transform \cite{park2017r}, k-nearest neighbors\cite{he2017novel}, Hilbert transform\cite{benitez2001use}, deep learning approaches\cite{laitala2020robust}\cite{vijayarangan2020rpnet}.

Photoplethysmography (PPG), commonly found in wearable devices, is a popular non-invasive method that measures blood volume changes through a light source and a photodetector. While PPG is user-friendly, it still requires wearing a device, and its accuracy can be compromised due to poor sensor placement.

Ballistocardiography (BCG) is the only method that can monitor long-term HR and HRV noninvasively, without contact, and without privacy concerns. BCG measures the mechanical vibrations of the body caused by the ejection of blood with each heartbeat\cite{starr1939studies}\cite{guidoboni2019cardiovascular}. This method can be integrated into everyday objects such as beds\cite{heise2011refinement} or chairs\cite{koivistoinen2004new}, providing continuous and unobtrusive monitoring, which is especially beneficial for long-term use. Recent studies have demonstrated the feasibility of using BCG for HR and HRV detection, generally relying on the detection of J-peaks in the signal. 
BCG has been shown to achieve accuracy comparable to ECG under certain conditions. The algorithm in \cite{lydon2015robust} uses the energy envelope of BCG signals and then locates the peaks of the envelope. The study reports an average heart rate error rate of 1.14\% for three young subjects and 3.7\% for four older adults over a one-minute time frame. K-means clustering\cite{rosales2012heartbeat} was implemented to classify the heartbeat. The study reports a correct detection rate ranging from 71.0\% to 92.5\% across four subjects. Adaptive thresholding has been introduced in \cite{choe2017simplified} for detecting the J-peaks in the BCG signal; this algorithm demonstrated a relative accuracy of 98.29\% with a root mean square error (RMSE) of 1.83 beats per minute (bpm) for HR across seven subjects. Deep learning networks, such as U-net and long short-term memory (LSTM) \cite{9176687}\cite{9743213} have been leveraged for their capability to learn complex patterns in physiological signals, and such approaches have demonstrated precision as high as 98\%. 

Detecting heartbeats from BCG signals presents challenges, particularly when monitoring older individuals. Unlike young, healthy subjects, the BCG signals of older adults often exhibit greater variability and lower signal quality. The J-peaks in BCG signals are less dominant and more difficult to isolate in older patients due to factors such as reduced cardiac output and changes in body composition. Additionally, the presence of comorbid conditions in older individuals can introduce further complexities. These difficulties highlight the need for advanced signal processing techniques and robust algorithms that can effectively filter out noise and adapt to the unique characteristics of BCG signals in older populations.
Our research introduces a method that uses a bed sensor to capture BCG signals and detect heart beats. By leveraging the naturalistic setting of a person's bed, our approach provides a comfortable and unstructured environment for monitoring. Improving the quality of the BCG signal itself presents significant limitations. Instead of solely focusing on noise filtering and peak detection directly from the BCG signal, we propose a transformer network inspired by previous studies\cite{vaswani2017attention}\cite{chiu2020reconstructing}\cite{tang2023ppg2ecgps}\cite{wang2022rec}\cite{zhu2021learning}. This network transforms the BCG signal into a corresponding ECG signal. Given that our primary focus is on heart beat detection rather than reproducing the full diagnostic ECG waveform, the precise point-wise accuracy between the predicted ECG (Pred-ECG) and the ground truth ECG (GT-ECG) is less critical. Instead, we emphasize the accurate identification of R-peak locations in the Pred-ECG. HR and HRV are the primary parameters of interest.
Previous studies have implemented multilayer networks that utilize a large number of parameters and substantial training. In \cite{wang2022rec}, ECG and BCG data are segmented using neighboring R-peaks and J-peaks as reference points, pairing the ECG slices with BCG slices that have the highest temporal overlap. This approach requires the computation of both R-peaks and J-peaks. In \cite{zhu2021learning}, two segmentation schemes were implemented, both of which require the detection of R-peaks and additional calculations to determine the 1/3 cycle length to the left of the R-peak for segmentation. In contrast, our proposed network has a simpler architecture and employs random segmentation with a 5-second sliding window and overlapping shifts, thereby reducing the computational workload and simplifying the preprocessing step before training process.

\section{Method}

\subsection{The Hydraulic Bed Sensor}

The hydraulic bed sensor (HBS) system\cite{heise2011refinement} is composed of four hydraulic bed transducers positioned under the mattress, as shown in Fig. 1. Each transducer is equipped with a pressure sensor capable of converting pressure into voltage signals. These sensors are sensitive for capturing the BCG movements of the body's center of mass associated with each cardiac cycle, as well as the movements of the rib cage wall corresponding to the respiratory cycle.
\begin{figure}[!t]
\centerline{\includegraphics[width= 0.4\textwidth]{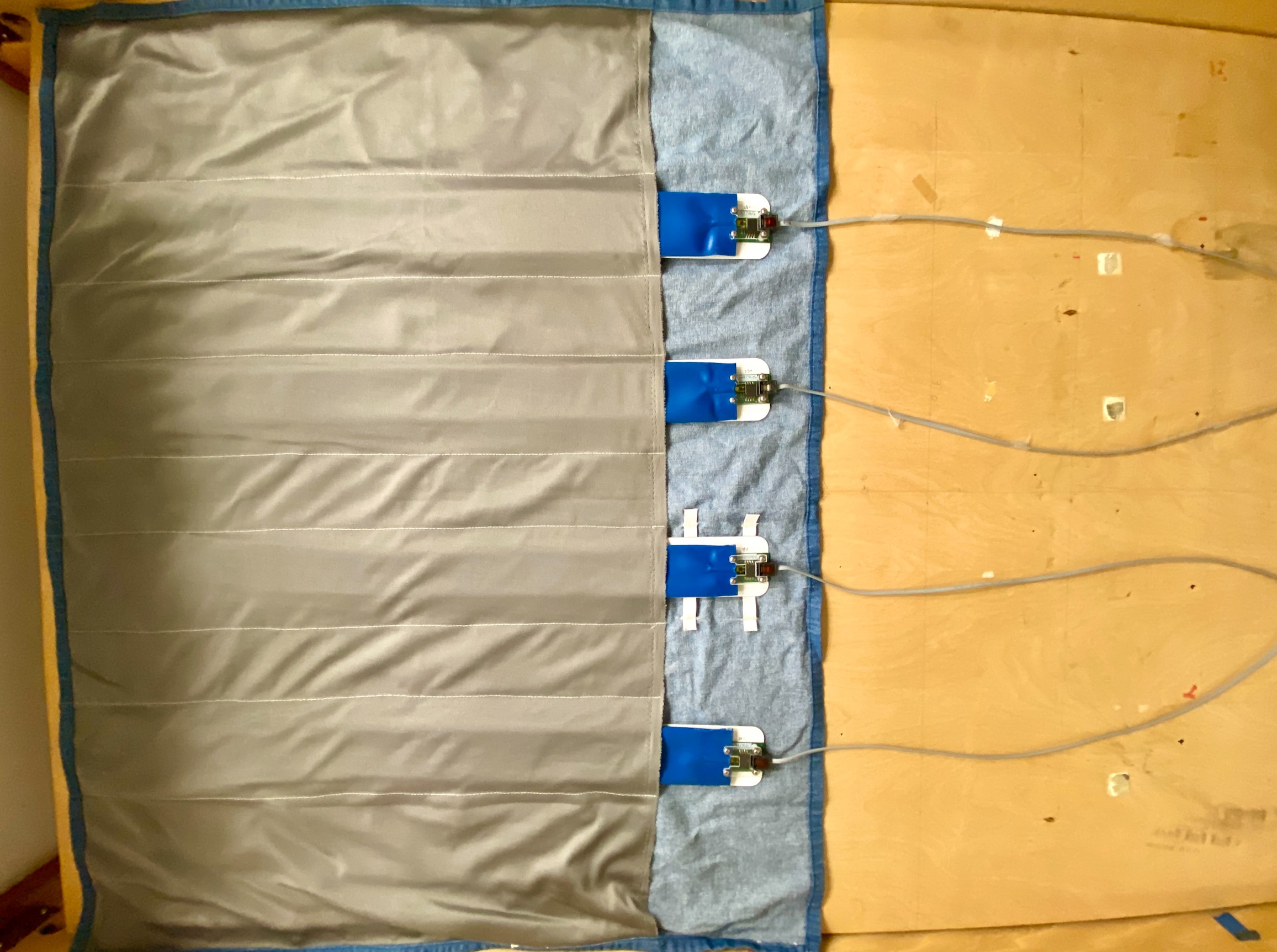}}
\caption{The hydraulic bed sensor system.}

\end{figure}

\subsection{Data Collection}

Three datasets were utilized in this study. The first dataset, referred to as the lab dataset, was collected from 46 young healthy volunteers (36 male, 10 female; $ 29.5 \pm 6.5$ years old) and the elder dataset was collected from 28 older residents (8 male, 20 female; $ 57.4 \pm 21.3$ years old) from a senior living facility. To ensure a comprehensive age range coverage, these two datasets were subsequently combined to form a third dataset, with 74 participants (44 male, 30 female; $ 40.0 \pm 19.5$ years old).

In both data collection settings, all subjects were positioned supine on the bed for a duration of 10 minutes. During this period, PPG, three-lead ECG, and BCG data were recorded simultaneously. The data acquisition was performed using an ADInstruments PowerLab 16/35 system, with the latest version of LabChart software. This study adheres to the guidelines and protocols approved by the University Institutional Review Board (IRB numbers: 2006391 and 2002586).

\subsection{Data Preprocessing}
The bed sensor data utilized in this study has eight output signals, including four raw signals and four hardware-filtered signals. To consolidate the data into a single-channel signal, an initial preprocessing step involves the selection of a single transducer based on the raw signals. For each sliding window size, the average amplitude of the raw signal was calculated for each transducer. The transducer with the highest average amplitude within each window was then selected for further analysis.
The original signals were collected at a sampling rate of 2000 Hz and subsequently downsampled to 100 Hz. A sixth-order Butterworth bandpass filter with a cutoff frequency range of 0.7-10 Hz were applied to remove high-frequency motion artifacts and low-frequency respiratory components. For segmentation, a sliding window of 5 seconds with a 0.25 second step was used to divide the 10 minutes signal into smaller segments. Both ECG and BCG signals were normalized between 0 and 1 for further processing.

\subsection{BCG to ECG Transformer module}
Fig. 2 illustrates the framework of the proposed network. The model architecture consists of an input linear layer, a positional encoding layer, a transformer encoder, and a fully connected output layer.
The input linear layer converts a 500 sample (\( 100 \, \text{Hz} \times 5 \, \text{seconds} \)) BCG signal into a dimension of 512. The positional encoding layer using sinusoidal functions captures the sequential nature of time series data and helps understand the temporal dynamics and dependencies in BCG signal. The transformer encoder is composed of 4 layers, in which each layer consists of 8 attention heads. The position-wise fully connected feed-forward network has an inner layer dimension of 2048. The self-attention mechanism represents as:
\begin{equation}
\text{Attention}(Q, K, V) = \text{softmax}\left(\frac{QK^T}{\sqrt{d_k}}\right) V
\end{equation}
where \( \mathit{Q} \), \( \mathit{K} \), \( \mathit{V} \) represents the query, key, and value with dimension $d_k$ which is 512. In this study we define h = 8 heads, The parallel multi-head self-attention layers is:
\begin{align}
    \text{Multihead}(Q, K, V) &= \text{Concat}(\text{head}_1, \ldots, \text{head}_h)W^O \tag{2} \\
    \text{head}_i &= \text{Attention}(QW_i^Q, KW_i^K, VW_i^V) \nonumber
\end{align}
where \(W^O\), \(W_i^Q\), \(W_i^K\), \(W_i^V\) are learnable weights.
At the end, a fully connected layer converts the matrix into a predicted ECG vector size of 500, which can be directly compared with the ground truth ECG. The training process utilizes Mean Squared Error (MSE) loss and the Adam optimizer. The model is trained 300 epochs with a learning rate of \( 1 \times 10^{-4} \) and batch size 8.

\begin{figure}[!t]
\centerline{\includegraphics[width= 0.3\textwidth]{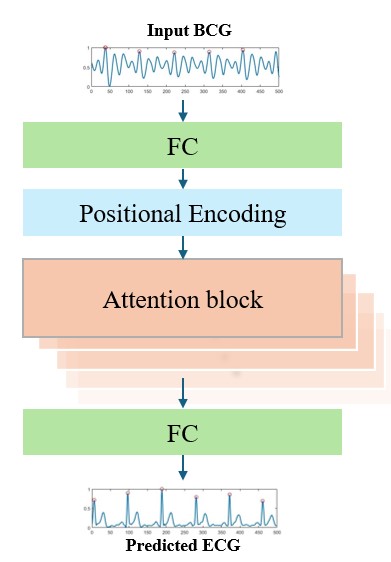}}
\caption{The framework of the proposed model. The model architecture consists of a fully connected layer (FC), a positional encoding layer, a transformer encoder, and an FC output layer.}
\end{figure}

\section{Experiments and Results}

In this study, we employed two different methods for training our model: the segment model and the subject model. 

\subsection{Segment Model}
The segment model was trained using three different datasets: the lab dataset, the elder dataset, and a combined dataset. For each dataset, all segments from the subjects were aggregated, shuffled, and then divided into five equal folds. In each iteration, one of the five folds was held out as the testing set, while the remaining four folds were combined to form the training set. This process was repeated five times, ensuring that each fold was used as the testing set once. The model’s performance was then evaluated on the testing set. There is no leakage of the testing set into the training set.

\subsection{Subject Model}
The subject model training involved spliting the data based on individual subjects. For the lab dataset, the subjects were divided into five folds with each fold including 9, 9, 9, 9, and 10 subjects respectively. For the elder dataset, each fold contained 6, 6, 6, 6, and 4 subjects respectively. In the combined dataset, 46 subjects from the lab dataset, and 28 subjects from the elder dataset were split into 5 folds, with each fold containing 15, 15, 15, 15, and 14 subjects. We ensured that each fold contained subjects from both datasets. In the fold with 15 subjects, 9 were from the lab dataset and 6 from the elder dataset; in the fold with 14 subjects, 10 were from the lab dataset and 4 from the elder dataset. The subject model allows for a more comprehensive evaluation of the model’s performance in a real-world scenario.

\subsection{Model Performance}

The data for ground truth ECG, predicted ECG, and input BCG were segmented into 5-second intervals. Within each segment, R-peaks in the ECG were identified using find\_peaks function, while the algorithm from\cite{lydon2015robust} was implemented as a baseline for heart beat detection directly from the BCG signal. Due to the BCG signal being a mechanical response to the electrical stimulus captured by the ECG signal, the BCG J-peak consistently lags behind the corresponding ECG R-peaks. This time delay is referred to as the R-J interval. Instead of directly comparing the beat locations between ECG and BCG, we focused on evaluating beat-to-beat HR and HRV. These metrics were derived from the reconstructed ECG and BCG signals and were compared to the ground truth ECG. To calculate the beat-to-beat HR within a 5-second segment, we use the median of the R-R intervals in that segment, with the R-R intervals converted into seconds. 
And use three HRV indices for each segment: Mean Heart Beat Interval (MHBI), root mean square of successive differences (RMSSD), and standard deviation of R-R intervals (SDNN). This approach provides a more comprehensive understanding of the physiological relevance and accuracy of the predicted signals. 
\begin{equation}
\text{HR} = \frac{60}{\text{median}\ RR_i }\tag{3}
\end{equation}

\begin{equation}
\text{MHBI} = \frac{1}{N} \sum_{i=1}^{N} RR_i\tag{4}
\end{equation}

\begin{equation}
\text{RMSSD} = \sqrt{\frac{1}{N-1} \sum_{i=1}^{N-1} (RR_{i+1} - RR_i)^2}\tag{5}
\end{equation}
\begin{equation}
\text{SDNN} = \sqrt{\frac{1}{N-1} \sum_{i=1}^{N} (RR_i - \overline{RR})^2}\tag{6}
\end{equation}
where \( RR_i \) is the \( i \)th R-R interval, and \( N \) is the total number of R-R intervals in segment. The HR is measured in bpm, while MHBI, RMSSD, and SDNN are measured in milliseconds (ms).

\begin{figure}[!t]
\centerline{\includegraphics[width=\columnwidth]{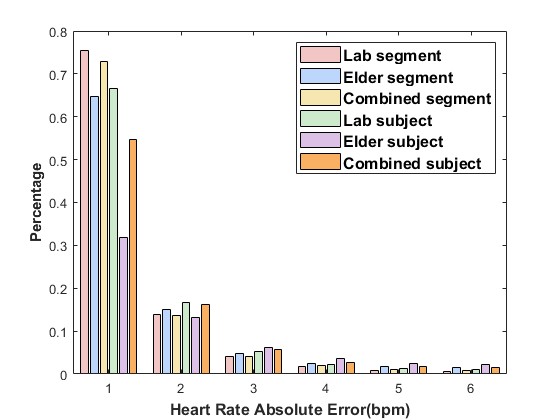}}
\caption{Histogram of the heart rate absolute error (bpm) distribution by comparing predicted ECG with ground truth ECG across all datasets and experimental models.}
\label{fig5}
\end{figure}

\begin{figure}[!t]
\centerline{\includegraphics[width= \columnwidth]{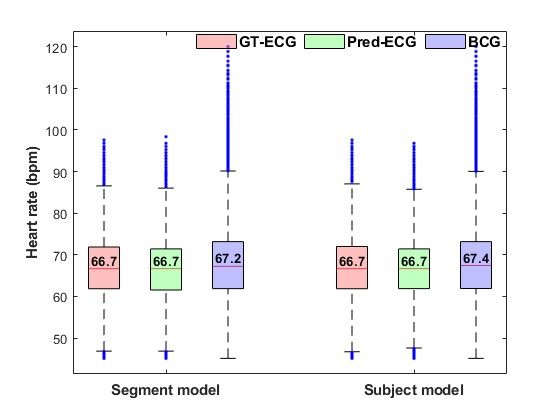}}
\caption{Boxplot of heart rate (bpm) estimation from ground truth ECG, proposed method from predicted ECG, and Lydon et al.[17] directly from BCG using the lab dataset.}
\label{fig1}
\end{figure}

\begin{figure}[htbp]
    \centering
    \begin{subfigure}[b]{0.9\columnwidth}
        \centering
        \includegraphics[width=\textwidth]{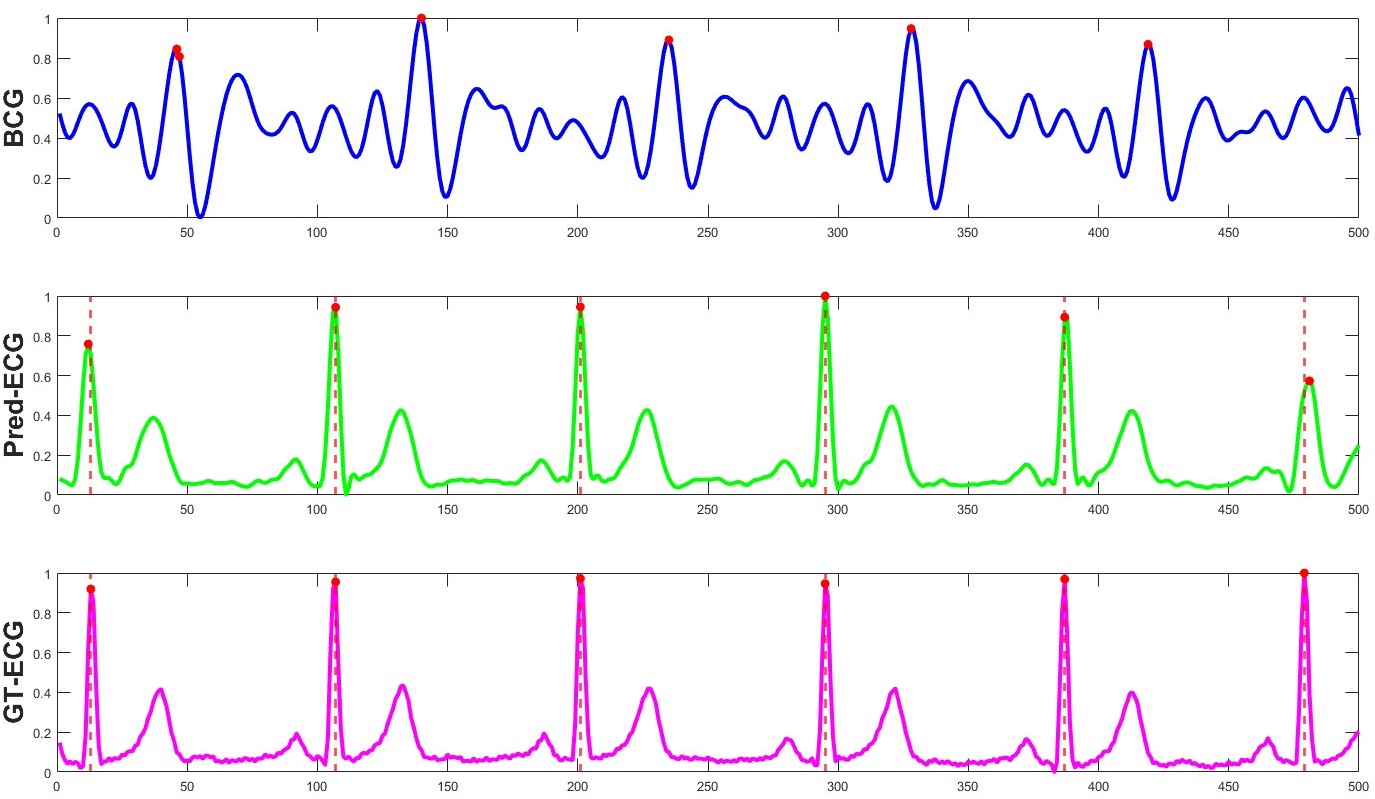}
        \caption{BCG with dominant J-peak}
        \label{fig:first_image}
    \end{subfigure}
    \vfill
    \begin{subfigure}[b]{0.9\columnwidth}
        \centering
        \includegraphics[width=\textwidth]{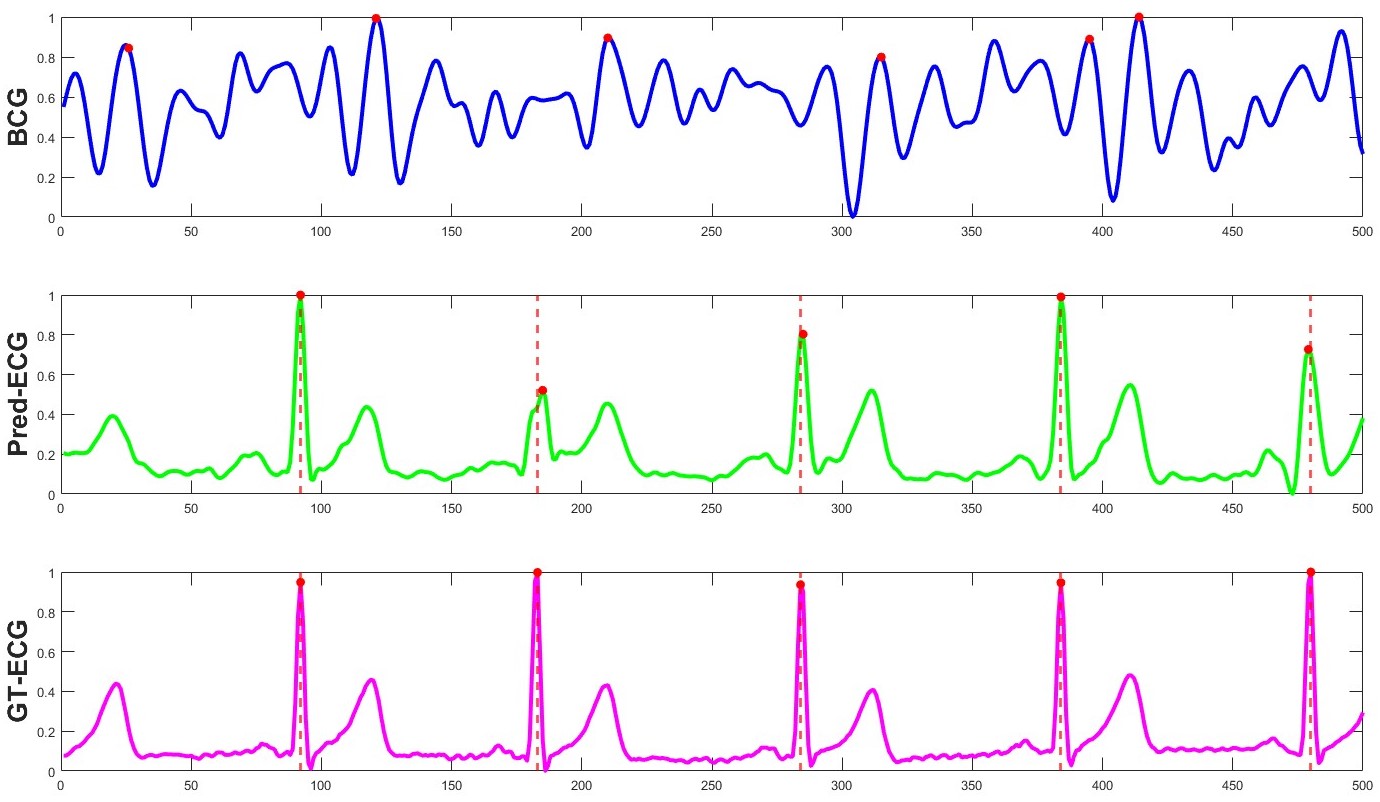}
        \caption{BCG without dominant J-peak}
        \label{fig:second_image}
    \end{subfigure}
    \caption{Comparison of BCG signal, predicted ECG from segment model, and ground truth ECG of a lab dataset participant. The red dots are the heart beats detected using Lydon et al.[17] from BCG, proposed method from predicted ECG, and R-peaks in the GT-ECG. The red dashed line superimposed on the Pred-ECG and GT-ECG subplots indicates where ground truth heart beats were detected in GT-ECG.}
    \label{fig:two_images}
\end{figure}

\begin{figure}[!t]
\centerline{\includegraphics[width= \columnwidth]{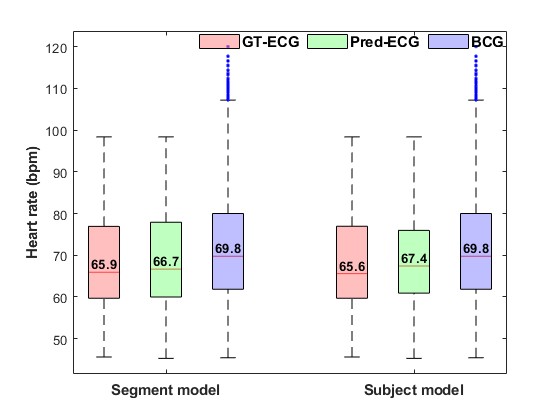}}
\caption{Boxplot of heart rate (bpm) estimation from ground truth ECG, proposed method from predicted ECG, and Lydon et al.[17] directly from BCG using the elder dataset.}
\label{fig1}
\end{figure}

\begin{figure}[htbp]
    \centering
    \begin{subfigure}[b]{0.9\columnwidth}
        \centering
        \includegraphics[width=\textwidth]{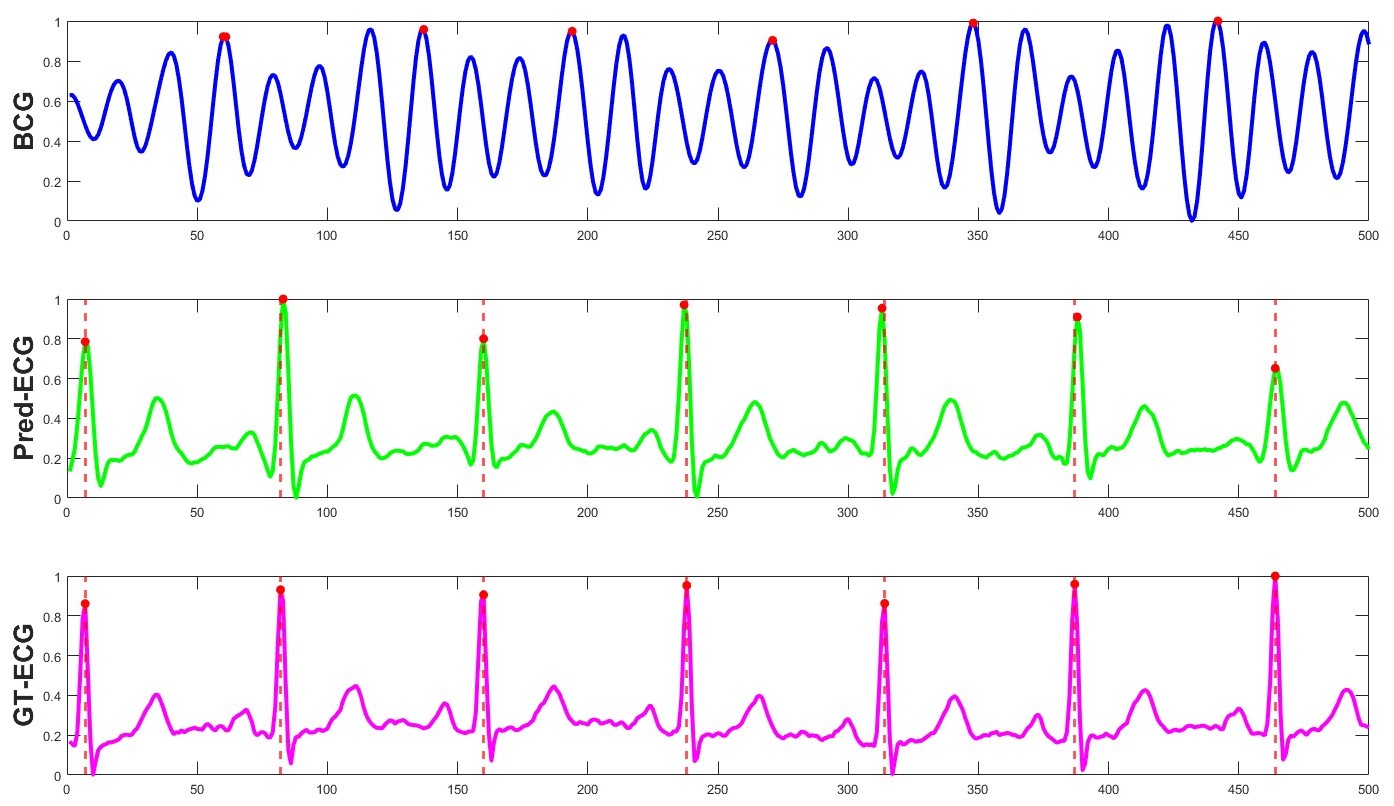}
        \caption{BCG with dominant J-peak}
        \label{fig:first_image}
    \end{subfigure}
    \vfill
    \begin{subfigure}[b]{0.9\columnwidth}
        \centering
        \includegraphics[width=\textwidth]{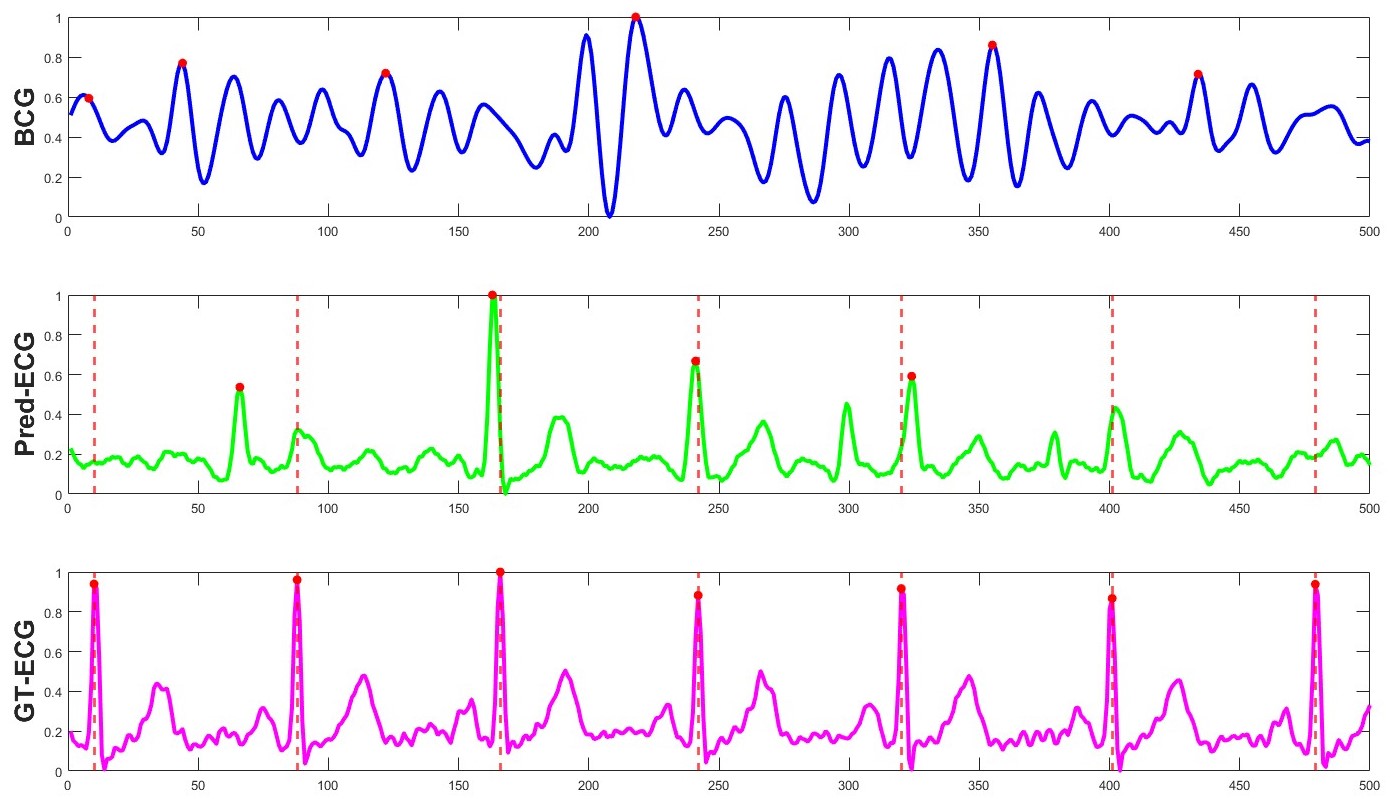}
        \caption{BCG without dominant J-peak}
        \label{fig:second_image}
    \end{subfigure}
    \caption{Comparison of BCG signal, predicted ECG from segment model, and ground truth ECG of an elder dataset participant. The red dots are the heart beats detected using Lydon et al.[17] from BCG, proposed method from predicted ECG, and R-
peaks in the GT-ECG. The red dashed line superimposed on the Pred-ECG and GT-ECG subplots indicates where ground truth heart beats were detected in GT-ECG.}
    \label{fig:two_images}
\end{figure}

\begin{figure}[!t]
\centerline{\includegraphics[width= \columnwidth]{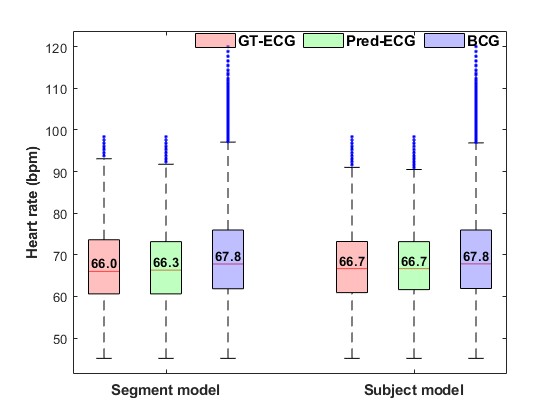}}
\caption{Boxplot of heart rate (bpm) estimation from ground truth ECG, proposed method from predicted ECG, and Lydon et al.[17] directly from BCG using the combination dataset (lab and elder participants together). The IQR of the combined dataset becomes narrower compared to using the elder dataset alone.}
\label{fig1}
\end{figure}

\begin{figure}[htbp]
    \centering
    \begin{subfigure}[b]{0.9\columnwidth}
        \centering
        \includegraphics[width=\textwidth]{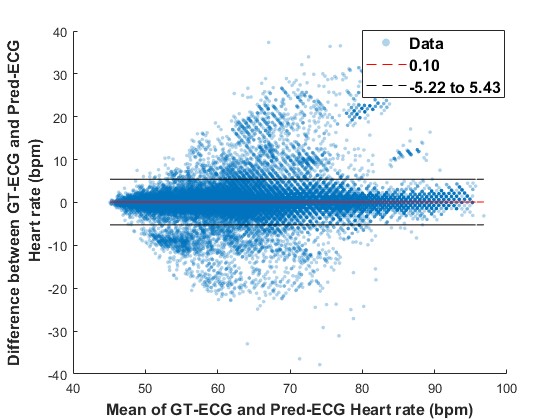}
        \caption{Lab segment model}
        \label{fig:first_image}
    \end{subfigure}
    \vfill
    \begin{subfigure}[b]{0.9\columnwidth}
        \centering
        \includegraphics[width=\textwidth]{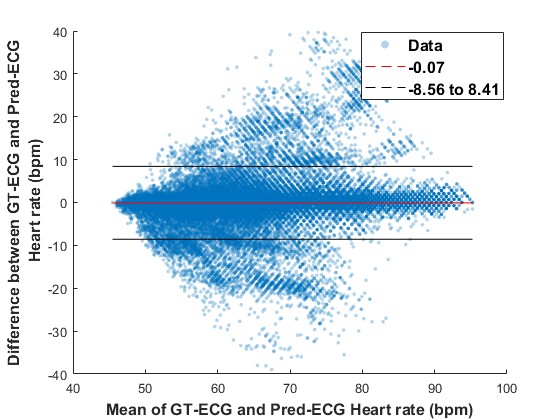}
        \caption{Lab subject model}
        \label{fig:second_image}
    \end{subfigure}
    \caption{Bland-Altman plot of HR of GT-ECG, Pred-ECG across two models on lab dataset.}
    \label{fig:two_images}
\end{figure}

\begin{figure}[htbp]
    \centering
    \begin{subfigure}[b]{0.9\columnwidth}
        \centering
        \includegraphics[width=\textwidth]{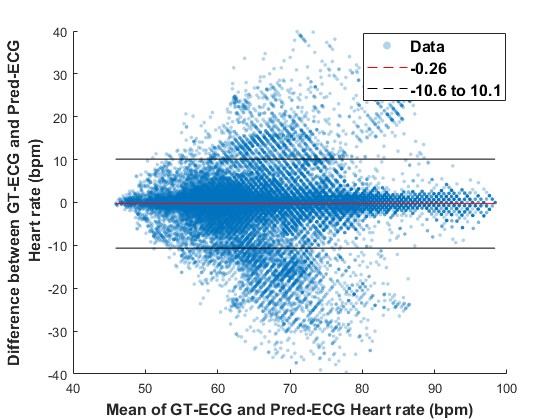}
        \caption{Elder segment model}
        \label{fig:first_image}
    \end{subfigure}
    \vfill
    \begin{subfigure}[b]{0.9\columnwidth}
        \centering
        \includegraphics[width=\textwidth]{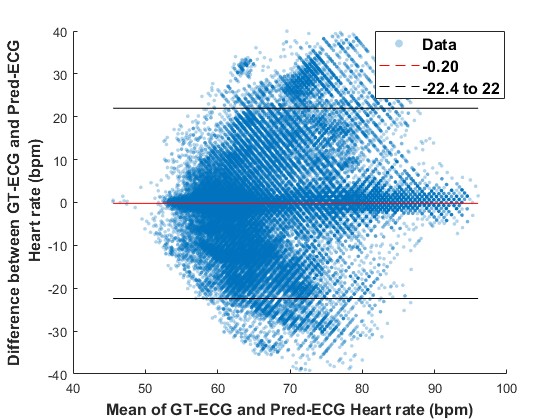}
        \caption{Elder subject model}
        \label{fig:second_image}
    \end{subfigure}
    \caption{Bland-Altman plot of HR of GT-ECG, Pred-ECG across two models on the elder dataset.}
    \label{fig:two_images}
\end{figure}

The histogram in Fig. 3 presents the distribution of heart rate absolute error (bpm) across different datasets and training methodologies. It indicates that the majority of the heart rate absolute errors are concentrated in the lower error ranges \mbox{(1-2 bpm)} for all categories.
The segment model consistently performs better than the subject model across all datasets. The elder dataset, when using the subject model, shows a deviation in performance; only 30\% of the segments fall within the 1 bpm error range. This reduced performance can be attributed to the fact that older individuals often have various cardiovascular conditions\cite{damluji2023management}, which affects the network's ability to generalize across subjects.

The boxplot of the lab dataset with the segment and subject models is shown in Fig. 4. The median heart rate is 66.7 bpm for GT-ECG and Pred-ECG, and slightly higher for BCG at 67.2 bpm and 67.4 for the two models. The inter quartile range (IQR) is consistent across GT-ECG and Pred-ECG but slightly wider across BCG, indicating more variability in HR measurement. The outliers and whiskers for BCG extend further compared to GT-ECG and Pred-ECG. The HR estimate from Pred-ECG is comparable to GT-ECG. Fig. 5 shows two example signals from the lab dataset. In Fig. 5(a), where the BCG signal has a dominant J-peak, the Pred-ECG closely matches the GT-ECG. Instead of point-wise matching, we concentrated on whether the predicted ECG could accurately indicate the peak locations in comparison to the R-peaks in GT-ECG. Interestingly, while the appearance of the R-peaks in the Pred-ECG in Fig. 5(b) differed from those of the GT-ECG due to the sub-optimal nature of the BCG, the heart beats were correctly detected. The alignment of detected heart beats between the Pred-ECG and GT-ECG is still consistent compared to the clear signal scenario in Fig. 5(a). 

The elder dataset heart rate estimation results are shown in Fig. 6. The median HR differences between the Pred-ECG and GT-ECG across the segment and subject models are 0.8 bpm and 1.8 bpm repectively. For BCG, the HR differeces are 3.9 bpm and 4.2 bpm. Although HR detection from Pred-ECG outperforms that from BCG, the overall performance is still lower than that observed with the lab dataset. The IQR is relatively wide compared to lab dataset. The elder dataset also display more outliers and extended whiskers, especially from the BCG signal, reflecting greater variability and challenges in heart rate prediction for the elder population. Combining the lab and elder datasets appears to reduce some of the negative effects seen in the elder dataset alone, as seen in Fig. 8. The HR predictions are more consistent, and the variance is reduced compared to using the elder dataset exclusively. This demonstrates the benefits of using combined datasets to enhance model performance and the potential improvement in handling diverse and challenging datasets involving older subjects.  

\begin{table*}
\centering
\renewcommand{\arraystretch}{1.5}
\label{table:1}
\caption{The Performance of Two Models on Three Datasets. Shows Total Number of Segment, Comparison of Pearson Correlation Coefficient of Heart Rate, MHBI, RMSSD, and SDNN Derived from Proposed Method using Predicted ECG and Lydon et al.\cite{lydon2015robust}  to Ground Truth ECG. The Best Results are Highlighted in Bold.}

\begin{tabular}{|P{0.9cm} P{0.9cm}|P{1.1cm}||P{0.9cm} P{1.6cm}||P{0.9cm} P{1.6cm}||P{1.0cm} P{1.7cm}||P{1.0cm} P{1.7cm}|}
\hline
 &  &  &\multicolumn{2}{c||}{HR (bpm)} & \multicolumn{2}{c||}{MHBI(ms)} & \multicolumn{2}{c||}{RMSSD(ms)} & \multicolumn{2}{c|}{SDNN(ms)}\\
&   & \makecell{Total \\ segments} & {Proposed}
& {Lydon et al.\cite{lydon2015robust}} & {Proposed}
& {Lydon et al.\cite{lydon2015robust}} & {Proposed}
& {Lydon et al.\cite{lydon2015robust}} & {Proposed}
& {Lydon et al.\cite{lydon2015robust}}\\
\hline

& Segments & 109203 & \textbf{0.97} & 0.80 & \textbf{0.97} & 0.84 & \textbf{0.66} & 0.48 & \textbf{0.69} & 0.50 \\

 Lab & Subjects & 108664 & 0.90 & 0.79 & 0.90 & 0.84 & 0.50 & 0.48 & 0.55 & 0.50 \\
\hline
 & Segments & 68797 & 0.92 & 0.62 & 0.91 & 0.66 & 0.50 &  0.31 & 0.51 & 0.31 \\

Elder & Subjects & 64989 & 0.47 & 0.59 & 0.39 & 0.64 & 0.34 & 0.47 & 0.35 & 0.51 \\
\hline
 & Segments & 171706 & 0.95 & 0.72 & 0.95 & 0.77 & 0.62 & 0.38 & 0.65 & 0.40 \\

Combine & Subjects & 174861 & 0.78 & 0.70 & 0.76 & 0.75 & 0.36 & 0.39 & 0.40 & 0.41\\
\hline
\end{tabular}
\end{table*}
Table I presents the overall performance of two models across three datasets. The correlation coefficients are compared between Pred-ECG and BCG against GT-ECG. The segment model consistently outperforms the subject model in all datasets and for all metrics. High Pearson correlation coefficients in the segment model indicate a strong agreement between the predicted ECG and ground truth ECG, particularly in the lab dataset. The elder dataset shows lower correlation coefficients, especially in the subject model, reflecting the challenges in accurately predicting HR and HRV metrics in elderly subjects. While the models perform well in estimating HR and MHBI, they struggle with capturing the variability in the data, as evidenced by lower RMSSD and SDNN correlations. This indicates the presence of outliers and variability in both datasets.
The Bland-Altman plots in Fig. 9 and Fig. 10 give an insight into the details of variability and outliers of data. The plots show the presence of disagreement between heart rates from GT-ECG and Pred-ECG. The subject model's wider spread indicates that there are more extreme differences, contributing to the lower correlations for variability metrics. The segment model's tighter clustering around the mean difference line and narrower limits of agreement indicate more reliable and consistent predictions, explaining its relatively better performance in RMSSD and SDNN correlations.

\section{Conclusion}

The goal of this study was accurate detection of heart beats. We introduced an efficient transformer network that takes a BCG signal as input and outputs a predicted ECG signal that facilitates improved heart beat detection. This approach addresses the challenges of detecting peaks from BCG signals without dominant J-peaks. As a non-invasive method, it holds potential for long-term HR and HRV monitoring, facilitating recognition of pattern changes associated health status changes thus enabling earlier intervention to mitigate deleterious effects of such changes.

The segment model demonstrates robustness across different datasets, maintaining higher correlation coefficients and narrower limits of agreement. It shows better performance in capturing both average HR and HRV. The subject model, which operates in a more realistic setting, shows comparable results on the lab dataset. The elder dataset highlights the challenges in predicting HRV in elderly subjects. The higher variance and presence of outliers indicate a need for improved models. Integrating lab and elder datasets improves model performance, suggesting that diverse data sources can enhance model robustness and generalizability.


\end{document}